# Cryptography Vulnerabilities on HackerOne

Mohammadreza Hazhirpasand
University of Bern
Bern, Switzerland
mohammadreza.hazhirpasand@inf.unibe.ch

Mohammad Ghafari
University of Auckland
Auckland, New Zealand
m.ghafari@auckland.ac.nz

*Abstract*—Previous studies have shown that cryptography is hard for developers to use and misusing cryptography leads to severe security vulnerabilities. We studied relevant vulnerability reports on the HackerOne bug bounty platform to understand what types of cryptography vulnerabilities exist in the wild. We extracted eight themes of vulnerabilities from the vulnerability reports and discussed their real-world implications and mitigation strategies. We hope that our findings alert developers, familiarize them with the dire consequences of cryptography misuses, and support them in avoiding such mistakes.

## I. Introduction

Cryptography is an essential part of software development. However, fatal mistakes by those who design cryptography APIs as well as those who use them can inflict undesirable outcome on software systems. Unfortunately, security issues in this domain can be devastating: in 2008, a vulnerability in the Debian OpenSSL package was discovered in which the generation of cryptographic keys had a limited entropy, resulting in reducing the space of possible keys to several hundred thousand [1]. More severely, in April 2014, the Heartbleed vulnerability allowed attackers to remotely read protected memory from an estimated 24–55% of popular sites powered by OpenSSL [1], or via POODLE attack, adversaries could downgrade and break the cryptographic security of SSL 3.0 and steal secure HTTP cookies [2].

Several work have focused on identifying crypto misuses. For instance, Lazar *et al.* found that 83% of the bugs occur due to misusing cryptographic APIs of various libraries and 17% of the bugs belong to the cryptographic libraries [3]. More recently, Hazhirpasand *et al.* found that 85% crypto APIs are misused in hundreds of Java open-source projects [4]. Nevertheless, we noted that the state of the art in crypto analysis often relies on static analysis tools, which are often prone to false alarms and limits the reliability of their findings [5] [6]. Moreover, to the best of our knowledge, previous work has barely discussed the consequence of such misuses.

In order to shed light on real-world crypto vulnerabilities and misuses, we ask the following research question : *"what crypto vulnerabilities security experts discover in practice?"*. To achieve this goal, we chose the HackerOne bug bounty platform as it hosts thousands of real-world, impactful security flaws that have been reported to companies. At the time of conducting this study, there were 408 seemingly crypto-related reports on HackerOne. Of which, our manual analysis revealed that 173 reports contain sufficient information for a thorough inspection of the reports. We found eight main themes of crypto vulnerabilities in which the SSL-related attacks theme was observed 58 times, and weak crypto defaults repeated 25 times. We discuss the vulnerabilities of each theme, their prevalence on HackerOne, and their mitigation strategies. Practitioners can benefit from our findings, which are based on real bug reports, to have a general overview of crypto vulnerabilities in practice, consequences, and remediations. It will be important that future research investigate other types of vulnerabilities, *e.g.,* authentication or authorization flaws, in order to create a more holistic view of related domains in practice. To carry out in-depth analyses of different types of vulnerabilities, we share the data extracted from HackerOne, containing 9311 disclosed reports, with interested researchers.[2]

The remainder of this paper is structured as follows. In section II, we explain why we chose HackerOne for this study. We present our methodology in section III and discuss our findings in section IV. We point out potential threats to the validity of this work in section V. We discuss related work in section VI. We conclude the paper in section VII.

## II. HackerOne

Bug bounty platforms have inspired security experts by drivers such as money, reputation points, or a place in the hall of fame [7]. In effect, security professionals are increasingly motivated to participate in such programs rather than requesting for CVEs alone.

Various bug bounty platforms exist such as HackerOne[3], Bugcrowd[4], Wooyun[5], and Synack[6], wherein highly trusted security researchers discover security vulnerabilities in companies' bounty programs. Although some studies analyzed the Wooyun platform, the website has not been accessible since July 20, 2016 [8].[7] HackerOne, Bugcrowd, and Synack are among the three most searched keywords, respectively, for bug bounty platforms on Google. [8] Despite the advantages of such platforms, we could only consider the ones in which there is a tendency to disclose vulnerability reports and the reports must entail adequate explanations. We conducted a preliminary check on HackerOne, Synack, and Bugcrowd to observe which platforms present detailed vulnerability reports. Bugcrowd enables access to bug bounty programs, detailed user statistics, and the number of vulnerabilities discovered and the average payout in each program. However, of 257 bug bounty programs, there were only 17 programs that had limited disclosed reports (*i.e.,* 54 reports). Similarly, the Synack platform also does not provide access to the detailed report of discovered vulnerabilities. In contrast to the previous two platforms, HackerOne permits hackers as well as organizations to study a vast number of preceding vulnerabilities' reports that are marked as disclosed. We, therefore, decided to study HackerOne.

---

[1] https://cve.mitre.org/cgi-bin/cvename.cgi?name=CVE-2008-0166
[2] http://crypto-explorer.com/hackerone/data.sql
[3] http://hackerone.com
[4] https://www.bugcrowd.com/
[5] http://www.wooyun.org/
[6] https://www.synack.com/
[7] https://en.wikipedia.org/wiki/WooYun
[8] Top bug bounty platforms searched on Google

TABLE I
SEVERITY LEVELS AND THE TOTAL AWARD GIVEN IN EACH LEVEL

| Severity level | Number of reports | Total award |
| --- | --- | --- |
| critical | 560 | $1 150 937 |
| high | 1205 | $1 433 357.7 |
| medium | 1448 | $293 957.96 |
| low | 2029 | $915 505.9 |
| none | 462 | $19 558.2 |
| null | 3607 | $1 529 183.25 |

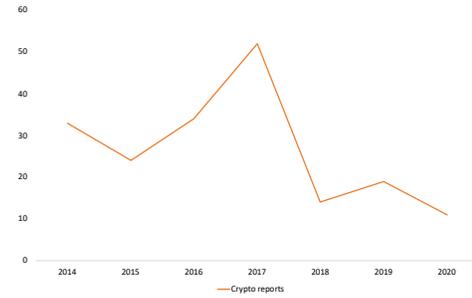

Fig. 1. The total number crypto reports per year

HackerOne was founded in 2012 and the company's network has paid more than $100 million bounties. On HackerOne, security teams, who are responsible for addressing security flaws discovered in a product, publish a program policy for a particular product. Security experts should carefully read the program policy before any report submission. In case of finding a security flaw, security expert should submit a report containing a comprehensive description of the discovery with clear, short reproducible steps. The report accordingly will be updated with the following events, when the vulnerability has been validated, if more information is needed from the security expert, or when the security expert is qualified for a bounty. The report is accessible to the security team immediately and remains private until the security team publishes a security patch. Public disclosure of a report can be requested by either the security expert or the security team.

We scraped HackerOne and there are 9311 disclosed vulnerability reports from 2013 to 2020, which is the time we conducted the scraping process. Of which, there are 560 reports designated as Critical and 1205 are specified as High severity. The vulnerability reports belong to 120 weakness types. Information disclosure has the highest number of vulnerability reports, *i.e.,* 979, and in total received $425 466 awards. Following that, the denial of service attack reports received $375 423 awards with 346 vulnerability reports. Interestingly, there are 3607 reports marked with Null severity, *i.e.,* without any rating, and 462 reports have None severity. However, submitting a severity rating for a report is optional for security experts; therefore, if the security expert does not specify the severity, it is assigned to be null.

Table I demonstrates how much money is awarded in each severity level. Interestingly, the vulnerability reports in the Null and None severity levels have received approximately $1,548,741 showing that not all vulnerabilities received the relevant severity level. For instance, the report id 112156 is set to no rating, *i.e.,* null, and the security researcher received $1000 due to a successful server-side request forgery (SSRF) attack. Each report has a state that increases or decreases the reporter's reputation. For instance, if a report's state is indicated as spam, the reporter receives -10 reputation. Overall, there are 224 duplicated, 870 informative, 215 not-applicable, 7978 resolved, and 24 spam reports in the dataset.

Among the top 25 weaknesses, the *"cryptographic issues - generic"* weakness type is at 17th place with 161 reports and $41 486 given award (See Figure 2). The same weakness type has 22nd place in terms of the amount of awards given to security experts. We also looked into the companies that had given awards to security experts for finding vulnerabilities in the *"cryptographic issues -generic"* weakness type. In total, the top 20 companies paid more than $40 000 to security experts. OpenSSL with 7 reports and $12 500 given award is in the first place, and Twitter with 12 reports and $2 940 has the highest number of reports. HackerOne also received 8 reports and paid $500 award.

TABLE II
THE STRUCTURE OF THE COLLECTED DATA FROM HACKERONE

| Field | Description |
| --- | --- |
| Title | the report's title |
| URL | the url of the report |
| To_whom | the company to which report is sent |
| By_whom | the person who sent the report |
| Username | the username of the sender of report |
| User_profile_pic | the user's profile picture |
| Severity_rating | the severity of the report (low/medium/high) |
| Severity_score | the score of severity (0 $\sim$10) |
| User_reputation | the reputation of the user |
| User_rank | the rank of the user |
| User_signal | the average reputation per report |
| User_percentile | the user signal percentile relative to others |
| User_impact | the average reputation per bounty |
| User_imp_percentile | the user impact percentile relative to others |
| Report_state | the state of the report (duplicate, informative, not-applicable, resolved, spam) |
| Report_date | the date that the report was submitted |
| Disclose_date | the date that the report was disclosed |
| Weakness | The weakness type assigned to the report |
| Award | The amount of money given to the reporter |
| Summary | The description describing how the vulnerability is discovered |

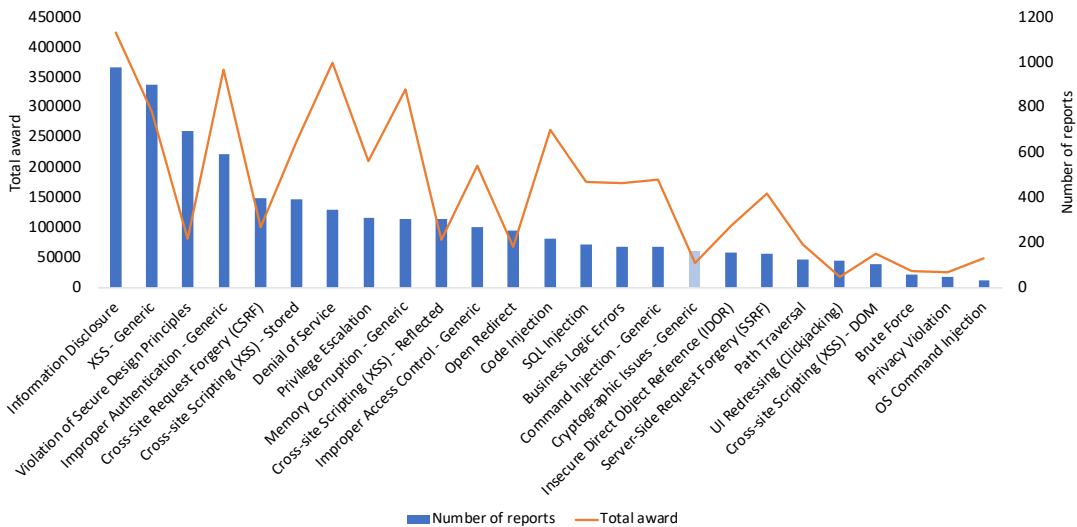

Fig. 2. Top 25 weaknesses and the total given award in each weakness

## III. METHODOLOGY

This section explains the structure of the collected data from HackerOne and the approach used to analyze crypto-related vulnerabilities.

### A. Schema

Recent bug reports are accessible on the hacktivity page[9] and search filters allow navigating disclosed reports. A more detailed inspection shows that HackerOne employs GraphQL query language to retrieve data from remote APIs. The data.hacktivity_items.pageInfo.endCursor property in the retrieved JSON is a key that can be used for retrieving the next page's content. We wrote a python script to automatically fetch disclosed reports from two APIs on HackerOne. The accumulated data are publicly available.[10]

The collected data contain 19 fields (See Table II). The title and URL fields are about the headline and the full web address of the vulnerability report. The to_whom and by_whom fields describe the company that receives the vulnerability report and the person who found the vulnerability. Regarding the vulnerability, the chosen severity rating and score are stored in the severity_rating and severity_score fields. The user_reputation field describes the points received or lost based on report validity. The user_rank field explains user ranking based on the earned reputation. The user_signal field stores information for identifying experts who have had consistently valid reports. The user_impact describes the expert's activity in terms of severity level. The user_percentile and user_imp_percentile fields help the security experts to compare their percentile signal and impact rank to other experts on the platform. The report_state shows the decision made for the report. The dates for when the report is submitted and disclosed are specified in the report_date and disclose_date fields. The weakness field reports the vulnerability type of the report and is our field of interest for finding cryptography-related vulnerabilities. The award field specifies the amount of money given to security researchers, and the summary field reflects the explanation of security experts for the identified vulnerability.

[9]https://hackerone.com/hacktivity
[10]http://crypto-explorer.com/hackerone/data.sql

TABLE III
THE SELECTED WEAKNESS TYPES AND THE ASSOCIATED NUMBER OF REPORTS

| Weakness | # reports |
|---|---|
| Cryptographic Issues - Generic | 161 |
| Weak Cryptography for Passwords | 7 |
| Use of a Broken or Risky Cryptographic Algorithm | 4 |
| Inadequate Encryption Strength | 3 |
| Missing Encryption of Sensitive Data | 3 |
| Missing Required Cryptographic Step | 3 |
| Use of Cryptographically Weak Pseudo-Random Number... | 3 |
| Use of Hard-coded Cryptographic Key | 2 |
| Reusing a Nonce, Key Pair in Encryption | 1 |

### B. Analysis

Our aim is to analyse vulnerability reports related to cryptography. Hence, the authors of this paper reviewed all the weaknesses, *i.e.,* 120 weaknesses, from the weakness field and extracted crypto-related keywords. Each weakness commonly consists of 3 or 4 words explaining the corresponding category of vulnerability, *e.g., Brute Force*, or *Command Injection - Generic*. After cross-checking the extracted keywords, they achieved a consensus on two keywords, namely "crypto" and "encrypt". Table III describes the selected weakness types and their associated number of reports in the data. The "Cryptographic Issues - Generic" type has the highest number of reports (*i.e.,* 161) and the "Reusing a Nonce, Key Pair in Encryption" type has the lowest number of reports (*i.e.,* 1) among the nine weakness types. In total, there are 187 vulnerability reports whose weakness type contained these two keywords. Of the 187 reports, there are 33 vulnerability reports in 2014 and there is a peak in 2017 with 52 vulnerability reports (See Figure 1). In 2018, 2019 and 2020, the number of crypto reports dropped to 14, 19, and 11, respectively. However, the data for 2020 may not be complete as we have not updated the data since the end of 2020 and more reports could be marked as disclosed. To increase the number of reports from other weakness types, we searched 40 crypto keywords, identified in the previous study [9], in the reports' summary. We found 221 unique reports containing at least one keyword in their summary.

Afterward, we used thematic analysis, a qualitative research method for finding themes in text [10], to find the frequent themes in the reports. We did not prepare a list of themes beforehand to assign the reports to the suitable themes. We derived the themes from finding patterns, commonalities over the course of the thematic analysis. The two authors of this paper individually read the selected reports (the title and summary fields), extracted the core problems of each report, built a list of themes, and assigned each report to a suitable theme. Each of them improved the extracted themes by reviewing the reports iteratively. Each reviewer stopped after three rounds of refinement and extraction since they noticed that the iterative process no longer adds anything of significance to the analysis. However, of 187 crypto-related reports, 33 reports did not have sufficient explanation regarding the found vulnerability, and accordingly, we omitted them. Of 221 reports whose summary contained crypto keywords, only 19 reports (*i.e.,* 8%) were related to cryptography. In general, we omitted the reports that were marked *not applicable* by the companies. Finally, the reviewers discussed the themes and associated reports with each other. We calculated Cohen's kappa, a commonly used measure of inter-rater agreement [11], between the two reviewers and obtained 71% Cohen's Kappa score, which manifests a substantial agreement between them. The reviewers re-analyzed the specific reports in a session where they had disagreements and achieved a consensus. They realized that their wording mechanism of themes differed in some cases and thus, unified themes were devised.

## IV. CRYPTO VULNERABILITIES

We present each theme of vulnerabilities, their prevalence in HackerOne's reports, and explain mitigation strategies suggested by the literature.

### A. SSL-related attacks

*POODLE:* The POODLE vulnerability can expose a man-in-the-middle possibility for attackers when using SSL 3.0 or under some circumstances for the TLS 1.0 - 1.2 protocols. Attackers are able to make 256 SSL 3.0 requests to reveal one byte of encrypted data. *Mitigation:* In order to mitigate the POODLE attack, SSL 3.0 must be disabled [2]. If disabling SSL 3.0 is not practical due to different reasons, the use of the TLS_FALLBACK_SCSV cipher suite ensures that SSL 3.0 is used only when a legacy implementation is involved, and thus, attackers are not capable of forcing a protocol downgrade. [2].[11]

[11]https://tools.ietf.org/html/rfc7507

*Sweet32:* The Sweet32 vulnerability enables attackers to reveal small parts of an encrypted message produced by 64-bit block ciphers, such as Triple-DES and Blowfish, under limited circumstances for TLS, SSH, IPsec and OpenVPN protocols. *Mitigation:* changing the default ciphers, such as 3DES or Blowfish, avoiding legacy 64-bit block ciphers, and selecting a more secure cipher like AES approved by NIST [12].[12]

*DROWN:* The DROWN vulnerability affects the OpenSSL library, SSL, and TLS on servers wherein SSLv2 connections are allowed [13]. Attackers can passively decrypt collected TLS sessions when a server supports SSLv2 as a Bleichenbacher padding oracle. The DROWN exploitation entails a chosen-ciphertext attack in order to steal a session key for a TLS handshake. *Mitigation:* server administrators must upgrade OpenSSL to the latest version and disable SSLv2, *e.g.,* use the following command in Apache webserver: *SSLProtocol All -SSLv2*.[13]

*BREACH:* The BREACH vulnerability is a weakness in HTTPS when HTTP compression is used [14]. The attack is agnostic to the version of TLS/SSL and applicable to any cipher suite. The attacker can obtain information about secrets in a compressed and encrypted response by attacking the LZ77 compression. In practice, the attacker injects random guesses into HTTP requests and measures the size of the compressed and encrypted responses to collect the smallest response sizes, meaning that the random guess matches the secret. *Mitigation:* To make the attack infeasible, HTTP compression must be disabled [15]. There are other countermeasures such as masking the secret with a one-time random value with each request and resulting in producing a new secret every time or to monitor and enforce request rate-limiting policy to discern nefarious visitors from genuine visitors [15].

*SSL stripping:* In an SSL stripping scenario, attackers downgrade the interaction between the client and server into an unencrypted channel to orchestrate a man-in-the-middle attack. There are several ways such as creating a hotspot, conducting ARP spoofing, and DNS spoofing to lure victims into the wicked network [16]. *Mitigation:* The SSL pinning technique is about avoiding man-in-the-middle attacks by checking the server certificates with a pinned list of trustful certificates added by developers during the application development phase [17] [18]. Other effective approaches against SSL stripping include: the HSTS (HTTP Strict Transport Security) header [19], History Proxy [20], and static ARP table [21].

*Freak:* The Freak vulnerability empowers attackers to intercept secure HTTPs connections between clients and servers and persuade them to employ 'export-grade' cryptography which presents out-of-date encryption key lengths [22]. The exploitation entails downgrading the RSA key length to 512-bit export-grade length in a TLS connection. *Mitigation:* Upgrade OpenSSL and the EXPORT grade ciphers must be disabled on the client side.[14] Moreover, it is necessary to use Perfect Forward Secrecy (PFS) cipher suites for key exchange,*e.g.,* Diffie-Hellman(DH) or Elliptic Curve DH in the ephemeral mode [23].

[12]https://sweet32.info/
[13]https://www.openssl.org/blog/blog/2016/03/01/an-openssl-users-guide-to-drown/
[14]https://access.redhat.com/articles/1369543

*BEAST:* The BEAST vulnerability provides a man-in-the-middle opportunity for attackers so as to reveal information from an encrypted SSL/TLS 1.0 session. The attack phase necessitates an adaptive chosen plaintext attack with predictable initialization vectors (IVs) and a cipher block chaining mode (CBC) [24]. *Mitigation:* Enable TLS 1.1 or preferably 1.2 which employ random IV and can also switch to using RC4 instead of using block ciphers [15].

*Certificate mis-issue:* The Certification Authority Authorization (CAA) DNS resource record allows a domain owner to set which Certificate Authorities are permitted to issue certificates for the domain. In that case, other CAA-compliant certificate authorities should refuse to issue a certificate. *Mitigation:* server managers should use the CAA feature in order to prevent issuing certificates by other unauthorized CAs [25].

***On HackerOne:*** There are various type of attacks against SSL in the vulnerability reports. The highest number of references belong to the POODLE and SWEET32 attacks that appeared in 13 and 7 reports. The Breach and Drown attacks each appeared 4 times in reports and the SSL pinning and Beast attacks each discussed 3 times in reports. Other types of attacks, such as SSL stripping, freak attack, CBC cut and paste attack, invalid curve attack, divide-and-conquer session key recovery, appeared only once in all reports. There are 25 reports that are about certificate-related issues, such as the CAA record or validation of a certificate. One of the major issues that security experts found was the missing DNS Certification Authority Authorization (CAA) record. Failure in certificate validation in mobile apps and insecure enabled RC4 cipher suites are the other evident reasons in the reports.

### B. Weak crypto defaults

Cryptographic algorithms require various parameters for which developers should provide. If an insecure argument or a weak crypto hash function, *e.g.,* MD5, is used, it may pose a severe threat to the security of a program. For instance, the use of insufficiently random numbers in a cryptography context leads to predictable values, or impersonating other users and access their sensitive information. Moreover, using a weak key length can also weaken the security of crypto algorithms to withstand brute-force attacks. *Mitigation:* Testing is an effective way to ensure the implementation is able to pass the basic security tests, *e.g.,* the National Institute of Standards and Technology (NIST) offers test vectors for a number of cryptographic primitives.[15] Moreover, NIST provides cryptographic standards and guidelines which is a reliable source for finding the right parameters and algorithms for various circumstances in cryptographic scenarios.[16] Employing static analysis tools in order to check crypto misuses in code snippets can immensely help developers in detecting crypto API misuses in the development phase, *e.g.,* CryptoLint [26] and MalloDroid [27].

***On HackerOne:*** There are 25 reports that using weak crypto defaults was one of the main issues. Experts found that weak encryption, *e.g.,* 512-bit RSA key or MD5, could lead to remote command execution, bypassing the download restriction for confidential files, decrypting data sent over SSH, gaining local file inclusion, and corrupting or modifying files on the server. There are 6 reports that security experts found bugs due to the existence of weak random number generators in systems. For instance, a bug could give access to the attacker to obtain access to OAuth access_token as a result of using PRNG instead of SecureRandom. In a report, a security researcher noticed that the target uses the TLS_RSA_WITH_3DES_EDE_CBC_SHA cipher, which is not marked as weak on SSL labs.[17] In four reports, the specified key size was problematic. A security researcher noticed that using a specific cryptographic function reduces the permutations of cryptographic keys. As a result, reduced permutations boost the chances of IV re-use. In two reports, the short length of keys, *e.g.,* 64 bit, increased the chance of finding key collisions and conducting brute force attacks.

### C. OpenSSL bugs

There are implementation bugs in crypto libraries that jeopardize the security of software systems using such libraries. What's worse, the software developers commonly cannot fix such vulnerabilities, and identifying them are beyond developer responsibility. Such flaws can be extremely dangerous as they expose thousands of software to security risks. *Mitigation:* Applying formal verification methods to examine the security properties of a cryptographic protocol can be considered, such as Cryptol [28] and PCL [29]. Furthermore, fuzzing approaches can also be part of library test suites or continuous integration in order to run before any versions are released, *e.g.,* the TLS-attacker framework is being used in MatrixSSL and Botan libraries [30].

***On HackerOne:*** Security experts found 25 bugs in the OpenSSL library and in total, they received $24 500 award. Security researchers reported vulnerabilities specifically to the implementation of OpenSSL. For instance, a security researcher reported that there is a mismatch in accepting a nonce value for the AEAD cipher. Other vulnerabilities concerned about consuming excessive resources or exhausting memory, recovering key on Diffie–Hellman small subgroups, performing padding oracle in AES-NI CBC MAC check, heap corruption, out-of-bounds read, and denial of service attacks.

### D. HTTP/HTTPs mixed content

The improper way of using HTTPS and mixing it with insecure network protocols, *e.g.,* HTTP or WebSocket (WS), can bring about undesirable outcomes for the website. Attackers can eavesdrop and conduct complicated attacks, *e.g.,* CRIME, JavaScript execution, cookie stealing, due to the unavoidable mistake of combining HTTPS and HTTP [31]. *Mitigation:* All the traffic should go through secure channels and the features that enforce secure connections on browsers, devices, and servers must be utilized. There are a number of HTTP headers that enforce and log the usage of HTTPS, namely Content-Security-Policy-Report-Only and Upgrade-Insecure-Requests.[18] The latter HTTP header necessitates that the browser must upgrade all insecure URLs before making any requests. For establishing secure WebSocket communications one must use the "wss://" instead "ws://" scheme, which utilizes port 443 [32].

---
[15]http://csrc.nist.gov/groups/STM/cavp/
[16]https://csrc.nist.gov/projects/cryptographic-standards-and-guidelines
[17]https://www.ssllabs.com/

*On HackerOne:* There are 22 reports in which security experts observed an insecure redirect from HTTPS to HTTP, lack of HTTPS on the login page, downloading resources from insecure channels, compromising HTTPS by loading resources from non-secure sources, downgrading from HTTPs to HTTP, and invitation reminder emails include HTTP links.

*E. Timing attacks*

Timing attacks exploit execution time differences. A timing attack is possible when a password check module immediately returns "false" as soon as the first character in the supplied password is different from the stored one. The attacker can observe the time it takes the system to respond to various queries until the correct password is extracted. *Mitigation:* In the aforementioned scenario, the developers should compare all of the characters before returning true or false. Returning an early response will leak information. Likewise, when the developers compare strings of equal length and drop the comparison when one string is longer or shorter causes information leakage about the secret string length. There are other preventive approaches, for instance, eliminating cache information leakage, including "random noise" into the computation, constant time implementations in the code, or using special APIs (*e.g.,* hmac.compare_digest in Python[19]) for checking hashed passwords [33] [34] [35].

*On HackerOne:* There are 11 reports that the timing attacks were the root cause of the vulnerability. Security researchers noticed that using == operator performs a byte-by-byte comparison of two values and once the two differ it terminates. In other words, the longer the process takes time, the more correct characters the attacker has guessed.

*F. Hard-coded secrets*

Hard-coded secrets can provide vital information for attackers in order to simply authorize themselves with privileged access in a software system. Such information can be commonly divided into three categories, namely, passwords/tokens, hard-coded usernames, and hard-coded private cryptographic keys. *Mitigation:* One advantage of static analysis tools is to prevent developers from using the same cryptographic key multiple times, hard-coded cryptographic keys for encryption or hashes [36]. The Common Weakness Enumeration (CWE-798) also suggests file system encryption techniques, secure architectural design, manual source code review, and dynamic analysis with manual results interpretation mitigation strategies.[20]

*On HackerOne:* There are 11 reports that the disclosure of secret keys or hard-coded passwords was the main theme of the reports. Security experts found secret keys or hard-coded passwords in different areas such as an error page, log files, or within a JavaScript source.

*G. HTTP header issues*

The attacker can eavesdrop on the HTTP connection to steal the victim's cookie or an XSS attack can retrieve the victim's cookie in case the cookie is not set properly. *Mitigation:* The "HttpOnly" flag prevents the access of the cookie from the client-side via JavaScript code in case of XSS exploitation. Moreover, to prevent hackers from eavesdropping on the cookies sent between client and server, the secure flag in cookies forces that the cookie must be only sent over an HTTPS connection [37]. The HSTS header resides on the browser, automatically redirects HTTP requests to HTTPS for the target domain, and does not permit a user to override the invalid certificate message on browsers [19].

*On HackerOne:* There are 9 reports that contained problems related to HTTP cookies and the HSTS security HTTP header. The main issue related to cookies is tied with not setting the SSL flag and HttpOnly in cookies. We also observed a lack of HSTS for websites and setting improper max-age for the header are frequent issues in the reports. Moreover, experts found that duplicate HSTS headers lead to ignoring HSTS on Firefox and it is feasible to downgrade a chosen victim from an HTTPS connection to HTTP.

*H. Miscellaneous attacks*

**KRACK:** The KRACK vulnerability concerns weaknesses in the WPA2 protocol four-way handshake for wireless networks. The exploitation process includes bypassing the official countermeasure of the 802.11 standard and reinstalling the group key, by combining WNM-Sleep frames with EAPOL-Key frames [38]. *Mitigation:* The users must update their Windows, OSX, Linux, Android, iOS, and firmware of home routers to address KRACK attacks [39].[21] The Wi-Fi Alliance should not solely test products for interoperability, but also employs fuzzing techniques to detect vulnerabilities [38].

**Hash length extension:** The hash length extension attack occurs if a developer prepends a secret value to a message, misuses a hashing algorithm in order to build a naive message authentication system [40]. In spite of the fact that the value of the prepended secret is confidential, if the attacker has knowledge about the string and the hash, he can still generate valid hashes. *Mitigation:* Developers must use the standard HMAC and avoid using MD5, SHA1, SHA-256, and other hashing algorithms [41].

**Chosen ciphertext:** The attack happens when the adversary can obtain the decrypted version of chosen ciphertexts. As a result, with the obtained pieces of information, the adversary can make attempts to recover the hidden secret key used for decryption [42]. *Mitigation:* To protect from such attacks, one can use secure crypto algorithms such as AES-GCM or RSA-PKCS#1 version 2 (RSA-OAEP) or consider the integrity and authenticity of the ciphertext [42] [43]. Developers must avoid using the RSA encryption standard PKCS #1 v1.

---

[18] https://www.w3.org/TR/upgrade-insecure-requests/
[19] https://docs.python.org/3/library/hmac.html#hmac.compare_digest
[20] https://cwe.mitre.org/data/definitions/798.html
[21] https://www.kaspersky.com/resource-center/definitions/krack

TABLE IV
THE EIGHT CRYPTO THEMES ON HACKERONE, UNDERLYING CAUSES, AND MITIGATIONS

| | # reports | Underlying causes | Remedies |
|---|---|---|---|
| SSL-related attacks | 58 | Using insecure SSL versions *e.g.*, 2.0, 3.0 and TLS 1.0/1.1 | Upgrade to more secure protocols (TLS 1.3) Upgrade OpenSSL to the latest version Avoid using 3DES, RC4,and Bluefish |
| Weak crypto defaults | 25 | Using wrong parameters and hashing algorithms (MD5, SHA-1) Short keys Insecure random number generators | Use stronger hashing algorithms *e.g.*, SHA-2 and SHA-3 Use guidelines to find the recommended key sizes Use cryptographic PRNG |
| OpenSSL bugs | 25 | Implementation flaws | Formal verification methods Fuzzing approaches |
| HTTP/HTTPs mixed content | 22 | Loading third-parties that have no SSL Having no measures to check the mixed-content | Install SSL certificate for the website Enforce HTTPS connections by the Upgrade-Insecure-Requests and Content-Security-Policy-Report-Only HTTP headers |
| Miscellaneous attacks | 12 | Using normal hashing algorithms (MD5) and prepending a message The KRACK attack Using early versions of RSA padding | The usage of authenticated encryption *e.g.*, AES-GCM or RSA-OAEP The usage of standard HMAC Upgrading router firmware |
| Timing attacks | 11 | Byte-by-byte comparison of two values | The usage of specific APIs to check two hashes (*e.g.*, constant_time_compare in Django) Including random-noise into computation |
| Hard-coded secrets | 11 | Placing static secret keys or hard-coded passwords in source codes or servers | File system encryption techniques Secure architectural design Manual source code review |
| HTTP header issues | 9 | Using cookies without the secure parameters Not enforcing the usage of HTTPS from client-side | The usage of the "HttpOnly" and "secure" flags The usage of the HSTS header |

***On HackerOne:*** In this category, we observed several attacks that were repeated in 12 reports. The padding oracle attack was repeated (*i.e.*, 4) more than other types, and encryption without authentication as well as insecure data storage each were repeated twice in the reports. The remaining attacks, *i.e.*, the hash extension length attack, the chosen-cipher text attack, the KRACK attack, and the Heartbleed attack appeared only once in the reports.

*I. Summary*

Table IV shows the eight crypto themes observed in these reports. The SSL-related attacks theme appeared more than other themes while the least appeared themes are OpenSSL bugs and HTTP header issues. We observe that the key reason for major issues is rooted in avoidance of using secure solutions by developers. For instance, despite the availability of TLS 1.3, there exist servers wherein SSL 3.0 is enabled. Developers commonly overlook the security parameters in cookies and this severely leads to downgrading the security levels of web applications. Such options are clearly elucidated in various reliable sources and even practical examples do exist. [22] [23] Furthermore, there exist encryption/decryption scenarios in which developers should comprehend several intricate concepts such as IV, differences between AES-ECB and AES-GCM, and key lengths. As as example, despite the inevitable consequence of deterministic random bit generators (DRBGs), developers paid inadequate scrutiny, leading to several fatal mistakes in software systems, *e.g.*, stealing $5 700 bitcoin due to a bug in entropy use in the Android DRBG. [24] Similarly, IV attacks also jeopardized the security of WEP protocol in wireless networks [44]. Some of the developers' mistakes can be prevented from occurring by employing security tools. For instance, the usage of outdated hashing algorithms or hard-coded secrets is easily identifiable with crypto misuse detectors. Nevertheless, there are various factors to dishearten developers from using such tools, *e.g.*, organizational and project-level constraints. Notably, the total number of participants in a study who use a static analysis tool is fewer than one-fifth (*i.e.*, 18) of the total number of participants (*i.e.*, 97) [45].

The themes are a summarized version of what has been improperly implemented in the industrial environment as well as signifying the seriousness of companies in addressing the issues. The provided real-world crypto vulnerability themes can assist developers, who lack knowledge in cryptography, to become more cautious while working with various elements, *e.g.*, OpenSSL, HTTP security headers, or hard-coded secrets, in the development phase. We also provided remedies, corroborated by previous research, for each theme to facilitate the hassle of finding secure practices. Future research should com-bine the current results with prevalent issues in other similar areas, *e.g.,* authentication, and authorization circumventions.

---

[22]https://developer.mozilla.org/en-US/docs/Web/HTTP/Cookies
[23]https://owasp.org/www-community/controls/SecureCookieAttribute
[24]https://arstechnica.com/information-technology/2013/08/google-confirms-critical-android-crypto-flaw-used-in-5700-bitcoin-heist/

## V. THREATS TO VALIDITY

To alleviate the errors of report categorization, each reviewer separately checked each bug report, and finally, they cross-checked their results. Besides checking the reports whose weakness field contained the "crypto" and "encrypt" terms, we also analyzed 221 reports whose summary contained 40 crypto-related keywords [9]. However, studying other reports (*i.e.,* 8903) on HackerOne may also uncover new relevant reports. In this work, our preliminary study revealed that Bugcrowd, and Synack do not provide detailed reports of discovered vulnerabilities, and hence, we only focused on vulnerabilities reported on HackerOne. We cannot lose sight of fact that the number of vulnerability reports in recent years has increased but many could have not yet been disclosed. On HackerOne, studying private or not yet disclosed reports is not possible. Furthermore, studying the available reports in detail is not feasible since the examined end-points or applications are not accessible to the public. Lastly, the analyzed vulnerabilities reports are associated with distinguished companies; and hence, such companies are more security conscious than the average software firms.

## VI. RELATED WORK

The study by Lazar *et al.* is the most relevant one to our work [3], wherein they performed a systematic study of 269 cryptographic vulnerabilities reported in the CVE database from 2011 to mid 2014. They categorized crypto vulnerabilities into four main groups, *i.e.,* plaintext disclosure, man-in-the-middle attacks, brute-force attacks, and side-channel attacks. Similarly, we also found a number of similar themes of vulnerabilities, *e.g.,* timing attacks or plaint text disclosure, showing that such issues are still relevant after seven years. Moreover, we pointed out crypto flaws found in bug bounty programs offered by well-known organizations and companies by which individuals can receive recognition and compensation and discussed preventive measures for each vulnerability in detail.

Braga *et al.* carried out a data mining technique, namely Apriori, to extract association rules among cryptographic bad practices, platform-specific issues, cryptographic programming tasks, and cryptography-related use cases from three popular forums: Oracle Java Cryptography, Google Android Developers, and Google Android Security Discussions [46]. They classified their findings into nine categories: namely Weak Cryptography (WC), Bad Randomness (BR), Coding and Implementation Bugs (CIB), Program Design Flaws (PDF), Improper Certificate Validation (ICV), Public-Key Cryptography (PKC) issues, Poor Key Management (PKM), Cryptography Architecture and Infrastructure (CAI) issues, and IV/Nonce Management (IVM) issues. In contrast to Braga's work, our work relates to real-world vulnerabilities that affected the security of organizations and companies.

Researchers have also developed new APIs to reduce the likelihood of crypto misuses. For example, Kafader and Ghafari developed FluentCrypto that hides the low-level complexities in using a native API and provides a task-based solution that developers can use without crypto knowledge [47]. It also allows crypto experts to configure the API as they find fit and determines that the configuration is secure at run time.

Research has extensively studied cryptography misuses in recent years. Chatzikonstantinou *et al.* conducted a study on how developers use cryptography in mobile applications [48]. They discovered that 87% of the applications had at least one crypto misuse in one of the four cryptographic categories defined in the study: usage of weak cryptography, weak implementations, weak keys, and weak cryptographic parameters. Hazhirpasand *et al.* analyzed hundreds of open-source Java projects that had used Crypto APIs and realized that 85% of Cryptography APIs are misused [4]. The authors contacted the developers and realized that security alerts in the documentation of crypto APIs are rare, developers may overlook misuses that originate in third-party code, and the code context must be considered; otherwise, it leads to the misunderstanding. The research community has proposed a number of tools to assist developers in securely using crypto APIs. For instance, the main aim of the CryptoExplorer platform, as a web platform, is to provide abundant examples for developers to study and follow the best practices [49].

The significance of bug bounty programs also cannot be overstated. Walshe *et al.* realized that the average cost of utilizing a bug bounty program for a year is less than the expenditure of employing two additional software engineers and such programs can be regarded as a complementary method to assist organisations in searching for security flaws [50]. Finifter *et al.*, reviewed the Google Chrome and Mozilla bug-bounty programs and observed that the aforementioned programs are more economical in comparison with employing a full-time security researcher [51]. Zhao *et al.* studied HackerOne and WooYun conducted a regression study that shows that monetary incentives have a significantly positive correlation with the number of vulnerabilities reported [8].

## VII. CONCLUSION

We were curious to observe what types of crypto flaws security experts find in real-world programs. We extracted disclosed vulnerability reports from HackerOne and analyzed the ones which were labeled as cryptography sensitive report. We found eight themes of crypto vulnerability in 173 reports. We introduced each theme's vulnerability, implications, prevalence on HackerOne, and suggest mitigation strategies. This study showed that developers do not commonly employ secure options in cryptography, leading to severe security vulnerabilities. Voracious readers and developers can learn from the findings, which are based on real bug reports, to avoid making the same fatal mistakes in practice.

VIII. ACKNOWLEDGMENTS

We gratefully acknowledge the financial support of the Swiss National Science Foundation for the project "Agile Software Assistance" (SNSF project No. 200020-181973, Feb. 1, 2019 - April 30, 2022). We also thank CHOOSE, the Swiss Group for Original and Outside-the-box Software Engineering of the Swiss Informatics Society, for its financial contribution to the presentation of this paper.